\documentclass[12pt]{article} \usepackage{amssymb,amsfonts,latexsym}
\usepackage[dvips]{graphics} \usepackage{epsfig}
 \normalsize

\def\N            {{\mathcal{N}}}
\def\R            {{\mathbb{R}}}
\def\C            {{\mathbb{C}}}
\def\Z            {{\mathbb{Z}}}

\begin{document}


\begin{flushright}  {~} \\[-1cm]
{\sf May 2005} \end{flushright}

\begin{center} \vskip 14mm
{\Large\bf Closed string tachyons in a smooth curved background} \\[20mm]
{\large Pedro Bordalo}
\\[8mm]
{\it Department of Mathematics,
King's College London}\\
{\it Strand, London WC2R 2LS, United Kingdom}\\
{\small bordalo@mth.kcl.ac.uk}
\end{center}
\vskip 18mm
\begin{quote}{\bf Abstract}\\[1mm]
Closed string tachyon condensation has been studied in orbifolds
$\C^2/\Z_{N,p}$ of flat space, using the chiral ring of the
underlying $\N=2$ conformal field theory.  Here we show
that similar phenomena occur in the curved smooth background
obtained by adding NS5-branes, such that chiral tachyons are localised
on lens submanifolds $SU(2)/\Z_{N,p}$.  We find a level-independent
subring which coincides with that of $\C^2/\Z_{N,p}$, corresponding
to condensation processes similar to those of hep-th/0111154.
We also study level-dependent chiral tachyons.

\end{quote}
\newpage


\tableofcontents

\section{Introduction and overview}

Time evolution between two metastable configurations in string
theory was related by A. Sen to a renormalisation group flow between
the underlying worldsheet theories.  The original proposal
\cite{Sen99} concerned dynamic processes of D-branes in a given
closed string background, driven by the better understood
condensation of open string tachyons.  In recent years, however, the
condensation of closed string tachyons has been addressed
ingeniously in a simple class of toy models, namely
non-supersymmetric orbifolds of flat space $\C^2/\Z_{N,p}$
\cite{APS01,HKMM01,Mar02}.

These toy models are examples of  theories with extended worldsheet
supersymmetry which nevertheless have space-time tachyons (and hence
break space-time supersymmetry).
The key idea is that this extended supersymmetry is preserved all
along the renormalisation group flows driven by target space tachyons which
are BPS states of the worldsheet theory.
This property allows to describe some properties of the endpoint of the flow;
thus, the flows studied for $\C^2/\Z_{N,p}$ were shown to describe a
gradual opening of the orbifold singularity, eventually leading in
the infrared to flat space.

One of the original difficulties of the topic -- that closed string
tachyon condensation would lead, according to the Zamolodchikov $c$-theorem,
to a decrease of the central charge -- has been turned into the so-called
$g_{cl}$-conjecture \cite{HKMM01}:  that {\it localised} closed
string tachyons in a non-compact space contribute at subleading
order (as compared to untwisted states) to the free energy of the theory,
and that it is this contribution $g_{cl}$
which decreases with their condensation.  In the toy models above,
the tachyons are localised at the orbifold singularity and
everywhere else the theory is free.

The purpose of this note is to show how this strategy can be applied
to study closed string tachyons in certain curved, smooth
backgrounds. These include orbifolds of the $SU(2)$ WZW model, the
generalised lens spaces $L_{N,p}=SU(2)/\Z_{N,p}$ studied in \cite{Bor03}.

The note is organised as follows: we first describe embeddings of
generalised lens spaces in string theory in section \ref{sec.embedd},
using NS5-branes.
In section \ref{sec.spectrum}, we briefly describe the corresponding CFT
constructions.  We show how space-time supersymmetry is broken and then
focus on the chiral ring of the $\N=2$ worldsheet superconformal algebra.
It turns out there exists a level-independent subring which is defined by the
pair $(N,p)$ and coincides with the chiral ring of the $\C^2/\Z_{N,p}$ theories.
Since the methods used in \cite{HKMM01} to analyse RG flows rely exclusively
on the structure of the chiral ring, such tachyons in our
backgrounds drive processes $(N,p) \rightarrow (N',p')$ just like
those described for $\C^2/\Z_{N,p}$. The level-dependent subring, on the
other hand, is identical to that of $\C/\Z_{k}$.  We save the conclusions
and outlook for section \ref{sec.outlook}.

\section{Smooth, curved, tachyonic backgrounds}
\label{sec.embedd}

Consider a stack of $k$ coincident NS5-branes in
$\R^{1,9}$.  It is well known that the 4-dimensional space transverse to the
NS5-branes becomes curved due to backreaction and in the near horizon limit
develops a throat of section $S^3$ whose radius is parametrized by the
dilaton:  the corresponding CFT is $SU(2)_k \times \R_{dil}$ \cite{CHS91}.
If instead we put the NS5-branes along the $\R_{1,5}$ directions of the
flat space orbifold $\R_{1,5} \times \C^2/\Z_{N,p}$, the transverse $SU(2)$
becomes a generalized lens space $L_{N,p}$.  To see this, let us consider the
$\Z_N$ actions involved.

If we parametrise $\C^2$ by the polar coordinates
$(r_1 e^{i\phi},r_2 e^{i\psi})$, then the $\Z_{N}$ acts as
$\phi \rightarrow \phi + 2\pi /N$ and $\psi \rightarrow \psi +  2\pi i p/N$.
Adding the NS5-branes then curves the transverse space to $SU(2) \times \R$,
where $SU(2)$ is spanned by $(\phi,\psi,\arctan (r_1/r_2))$, up to these
identifications.  If we now switch to the Euler coordinates
$g=e^{i\frac{ \psi+\phi }{2}\sigma_3} e^{i\frac{\theta }{2}\sigma_1}
e^{i\frac{\psi-\phi }{2}\sigma_3}$ for $SU(2)$, these identifications become
\begin{equation}\label{eq.action}
g \sim \omega^{\frac{p+1}{2}} g \omega^{\frac{p-1}{2}}
\end{equation}
where we chose $\omega=e^{i\sigma_3/2}$ to be the $\Z_{N}$ generator.
The quotient by (\ref{eq.action}) defines the lens space $L_{N,p}$.  
Notice that $p$
takes values in $\{1,\ldots,N\}$.  This action is free as long as
$p$ and $N$ are mutually prime, which we will assume in the
following.  The particular case $p=1$ corresponds to the more
familiar left quotient of $SU(2)$ (furthermore, $L_{N,p}$ and $L_{N,-p}$
are diffeomorphic, being related by a conjugation by $\omega$).
The various lens spaces with a given $N$ are topologically very
similar (they have the same cohomologies, but can be distinguished
by certain knot invariants \cite{Brown}).

Thus the near horizon limit of
our configuration is $L_{N,p} \times \R_{dil}$ as advertised.  Notice
in particular that adding the NS5-branes smoothens out the orbifold
singularity by replacing it with the infinite throat.  On the other hand,
the $\Z_N$ quotient in general breaks all spacetime supersymmetry, since
it acts differently on the right and on the left.  In fact, analysing the
spectrum in the next section, we will see that theories with $p\neq 1$ 
have spacetime tachyons.

\section{CFT description and chiral rings} \label{sec.spectrum}

Let us be more precise in the conformal field theory description of
our configuration of NS5-branes.
Adding $k$ NS5 branes at the origin of flat space yields
$SU(2)_k \times \R_{dil}$. The $SU(2)$ transverse to the NS5-branes
can now only be
orbifolded by an action whose order $N$ divides the level $k$, such that
the exponential of the Wess-Zumino action is well defined \cite{Bor03}.
Thus, we are in fact considering a system of $k=NN'$ NS5-branes spread out
in a circle at a finite distance of the orbifold singularity.
This singularity, which would describe the strong string coupling
region where the CFT description breaks down, is thus naturally
cut off from the theory.  According to the analysis 
of \cite{Sfet99}, the chiral algebra describing the theory near the 
NS5-branes is then $SL(2,\R)/U(1) \times SU(2)/U(1)$.  

To construct the superconformal field theory appearing in
this near horizon limit, let us
recall the bosonic CFT describing string propagation in lens spaces
$L_{N,p}$ \cite{Bor03}.  As explained in the
introduction, for our study of tachyon condensation, we need an
$\N=2$ supersymmetric and non-compact version of that.  The simplest
extension is to add one non-compact direction to the
three-dimensional $L_{N,p}$.\footnote{Recall that $\N=2$ worldsheet
supersymmetry implies a target space with complex structure, and
thus with even dimension.}  For that purpose, recall that the
relevant conformal field theory has as chiral algebra $SU(2)/U(1)_k
\times U(1)_k$, and that its spectrum is given by fields of the form
$\Phi_{j,m}^{pf}\Phi_n^{u(1)}\cdot
\bar{\Phi}_{j',m'}^{pf}\bar{\Phi}_{n'}^{u(1)}$ (where the $pf$ stands
for the parafermions $SU(2)/U(1)$) with selection rules
\begin{eqnarray}
m-n=(p+1)\frac{k}{N} r ,\ \ \ m+n=(p-1)\left( \frac{k}{N}r+s \right)
\mathrm{mod\ } 2N \label{eq.spectrumL}\\
m' = m+2s, \ \ \ \ \ \ \ \ n'=n+2s+2\frac{k}{N}r, \ \ \ \ \ \
\ \ \ \ j'=j \ \ \ \ \ \ \ \label{eq.spectrumR}
\end{eqnarray}
with $r=0,\ldots,N-1$ and $s=0,\ldots,k-1$ and $p \in \{1, ldots ,N\}$ 
(recall that $N$ divides
$k$).  In CFT terms the orbifold group is $\Z_k \times \Z_N$, where the
$\Z_N$ factor acts only on the $u(1)$ chiral algebra, so there 
are $Nk-1$ twisted (left-right asymmetric) sectors, corresponding 
to non-zero values of $r$ and $s$.  Notice that different choices of $p$ 
only affect the set of asymmetric fields in the spectrum; in (rational) 
conformal field theory this choice is referred to as discrete 
torsion \cite{KZ93}\cite{Va86}.

In moving from the bosonic $L_{N,p}$ towards an $\N=2$
superconformal field theory, it suffices to replace the $U(1)$
factor, since the parafermions $SU(2)_k/U(1)$ are naturally an
$\N=2$ minimal model.  Given the analysis above, the most natural 
choice is to start from the chiral algebra
\begin{equation}\label{algebra}
\mathcal{A}=\frac{SL(2,\R)_{k+2}}{U(1)} \times \frac{SU(2)_{k-2}}{U(1)}
\end{equation}
The level of the $\N=2$ ``cigar CFT'' $SL(2,\R)/U(1)$ has been chosen
such that the chiral algebra has (bosonic) central charge $c=4$. As
the name "cigar" suggests, the target space of the full $SL(2,\R)/U(1)$ CFT
is a semi-infinite cylinder whose compact direction
shrinks to zero size at the origin.  
The natural interpretation of a CFT with symmetry $\mathcal{A}_L\times
\mathcal{A}_R$ and spectrum (\ref{eq.spectrumL},\ref{eq.spectrumR}),
would be that of a lens space whose radius grows along the
non-compact cigar dimension. Indeed, different orbifolds of $\mathcal{A}$
have already been used to describe $AdS_3 \times SU(2)$, namely
\cite{GKS98}
\begin{equation}\label{algebra2}
\frac{\frac{SL(2,\R)}{U(1)} \times U(1)}{\Z} \times \frac{
\frac{SU(2)}{U(1)} \times U(1)}{\Z_k}
\end{equation}
and also to describe $\R_{\phi}\times SU(2)_k$, namely \cite{GK99}
\begin{equation} \label{algebra1}
\frac{SL(2,\R)/U(1) \times SU(2)/U(1)}{\Z_k}
\end{equation}
In both cases, the orbifold acts at the level of the chiral algebra, ie. it
is a symmetric orbifold. In contrast, generalising the $SU(2)$ factor in
these models to lens spaces involves imposing the asymmetric orbifold
constraints (\ref{eq.spectrumL},\ref{eq.spectrumR}) on the spectrum.

\bigskip

While the proof of modular invariance of general simple current partition
functions is complicated \cite{KZ93}, that for the $L_{N,p}$ case relies only in
the modular properties of the $U(1)$ characters, much like Vafa's original
analysis of discrete torsion in the 2-torus \cite{Va86}.  Nevertheless,
this analysis does not
carry straight away to our case here, because (unlike for the parafermions) the
modular transformations of $SL(2,\R)/U(1)$ extended characters do not
factorize
into $SL(2,\R)$ and $U(1)$.  Fortunately,  
in the following the explicit expression
of the partition function is not essential, so we will just assume that our
theory is defined by the triplet $(k,N,p)$ such that its bosonic
sector is given by (\ref{eq.spectrumL},\ref{eq.spectrumR}),
as justified by the geometric arguments of the previous section.

\bigskip

We can now nevertheless see how space-time supersymmetry is broken.
As mentioned in \cite{KLL98}, the left-moving supercharges are
mutually local with the fields of the theory if
the latter obey $$\frac{m-n}{k} \in \Z $$ since that is the
conformal weight appearing in the most singular term of the OPE of
the supercharges with a given state (and similarly for the right movers).
This means that only for $(N,p)=(1,1)$ or $(k,1)$ are all the
left- and right-moving supercharges conserved (actually,
the configurations $(N,1)$ and $(k/N,1)$ are T-dual to
each other \cite{Bor03}).  In this case
there is a GSO projection which leads to a consistent type II string
theory \cite{GKS98}.  For non-trivial quotients (both $N$ and $N'$ greater
than $1$) with
$p=1$, only the left-moving supercharges are conserved
\cite{KLL98}.  Again there exists a GSO projection (the restriction
of the one above to a chiral half of the Hilbert space) such that
these supercharges still generate a consistent type II superstring
theory\footnote{However,
in \cite{KLL98}, only the untwisted part of the spectrum of this
theory was studied since the purpose was to match it to the
supergravity description.}. Finally, when $p \neq 1$, none of the
supercharges is conserved and there is no space-time supersymmetry.

\subsection{The chiral ring}

Chiral fields $\Phi$  of the $\N=2$
superconformal algebra saturate the BPS bound
\begin{equation}\label{eq.chiral}
\Delta_\Phi=\frac12 Q_\Phi
\end{equation}
where $\Delta_\Phi$ is the conformal weight of $\Phi$ and $Q$ is its
R-charge.\footnote{The worldsheet $\N=2$ symmetry factorises
naturally in the algebra $\mathcal{A}$ (\ref{algebra}). Here we choose
fields which are chiral for both the cigar
and the parafermions, and also for both the left and the right-moving
algebras.  Other choices, involving antichiral fields with
$\Delta_\Phi=-\frac12 Q_\Phi$ would lead to similar results.}
Since $Q$ is additive, the set of chiral fields forms a ring, called the
chiral ring \cite{LVW88}.  Following \cite{HKMM01}, we are interested 
in the chiral ring as a way to characterize our theories $(k,N,p)$ which 
is manageable under RG flows. Indeed, one can see in a lagrangian
formalism that this BPS property is preserved along the renormalization 
group flow driven by a chiral relevant perturbation. The chiral ring 
is deformed under such a flow, but can provide information 
about the infrared fixed point.

Chiral fields are always in the NS sector.
The NS representations of our chiral algebra are specified by 4 indices,
eg. the left moving NS states are $|j,m\rangle^{cg} |l,n\rangle^{pf}$.
The parafermion indices are restricted to $l=0,\frac12,\ldots,k/2-1$ and $n$ is
an integer defined modulo $2k$ such that $2l+n =0$ mod $2$.  The cigar
representations (normalizable and unitary, with integer level $k+2$) are 
similarly labelled by two parameters $(j,m)$ and fall into three classes:
the positive discrete series, with $j=0,\frac12,\ldots$ and 
$m = j+t$ where $t\in \mathbb{N}$; the negative discrete series, with 
$j=0,\frac12,\ldots$ and $m\in -j-t$ where $t\in \mathbb{N}$; 
and the continuous 
representations, with $j=\frac12 +i\lambda$, where $\lambda \in \R$, and 
$m\in \Z$.  Their weights and charges are
\begin{eqnarray}
\Delta_{jm}^{cg} = \frac{m^2-j(j-1)}{k}, \ \ \ Q_{jm}^{cg}=-\frac{2m}{k}
\mathrm{mod\ } 2 \label{chargescg}\\
\Delta_{ln}^{pf} = -\frac{n^2}{4k} +\frac{l(l+1)}{k}, \ \ \
Q_{ln}^{pf} = -\frac{n}{2k} \ \mathrm{mod\ } 2\label{chargespf}
\end{eqnarray}
The chiral fields are therefore of the form
\begin{equation}\label{eq.chiralfields}
|j,-j\rangle^{cg}_L |l,l\rangle^{pf}_L \otimes
|j',-j'\rangle^{cg}_R |l',l'\rangle^{pf}_R
\end{equation}
Selecting the chiral fields (\ref{eq.chiralfields}) which obey the selection
rule (\ref{eq.spectrumL},\ref{eq.spectrumR}) will then yield the 
chiral fields in the theory with target space $L_{N,p}\times \R_\phi$, see below.

\medskip

Before we go on to study the structure of the chiral ring, notice that
only the discrete representations of $SL(2,\R)/U(1)$ appear in the
chiral fields.  Since these states are localised at the tip of the
cigar, the chiral states of (\ref{algebra}) will be localised
at the origin of the non-compact direction.  In particular, and even though
the target space is smooth, the chiral tachyons (those chiral fields
with $Q<1$, ie. chiral relevant operators) are {\it localised in
the $L_{N,p}$ submanifold}.  Rolling
down the potential of such tachyons will therefore not change the total
central charge, which is dominated by large volume contributions
\cite{APS01,HKMM01}.

\medskip

Let us now combine the $\N=2$ chirality condition (\ref{eq.chiral})
with the lens space selection rules (\ref{eq.spectrumL},\ref{eq.spectrumR}).
The analysis can be made at the CFT chiral level; left-moving chiral fields
satisfying (\ref{eq.spectrumL}) can be divided into three branches, specified
by their cigar and parafermionic $U(1)$ charges (previously called $m,n$):
\begin{eqnarray}
\mathrm{I: fields\ with\ } r\neq 0, s=0, & \mathrm{of\ the\ form\ } &
W_r = |-rp\frac{k}{N}\rangle^{cg}|r\frac{k}{N}\rangle^{pf} \nonumber \\
\mathrm{II: fields\ with\ } r= 0, s \neq0, & \mathrm{of\ the\ form\ } &
V_s = |(p-1)s\rangle^{cg}|(p-1)s\rangle^{pf} \label{eq.twistfields} \\
\mathrm{III: fields\ with\ } r=s=0, \ \ \ & \mathrm{of\ the\ form\ } &
T_c = |Nc\rangle^{cg}|Nc\rangle^{pf}
\end{eqnarray}
where in branches I and II $s=1,N-1$ and $r=1,k-1$ 
and in branch III $c=1,\ldots,2k/N$. 
Recall that fields with $r,s\neq 0$ have different left and right
$U(1)$ charges, via the selection rules (\ref{eq.spectrumR}). In particular,
since for lens spaces the parafermionic number $l$ must be the same in the
left-movers and the right-movers, $l=l'$, the chiral left-movers with a
non-zero $s$ value (those in branch II) are coupled to non-chiral right
movers and vice-versa. Similarly, in branch I the variable $r$ shifts
the $U(1)$ charge $n'$ associated to the right-moving $SL(2,\R)/U(1)$.
So branch I and II do not give rise to chiral fields of the full conformal
field theory.

Finally, the fields in branch III give rise to left-right symmetric (ie.
independent of $p$) chiral fields of the full CFT.  This branch includes
tachyonic fields for any $N>1$, even in the supersymmetric cases reviewed
in section \ref{sec.embedd} when $p=1$.  We are led to conclude that 
these fields are projected out in the supersymmetric $p=1$ case, and 
perhaps also for general $p$.  In any case, these fields are in the  
untwisted sector, and so we will not consider them in the following.

\subsection{Twist fields}

So we have no twisted chiral states in our spectrum.  Nevertheless, the
twisted sectors are just as well described by the twist fields
which permute them \cite{DFMS87} (even though they are not in the
partition function, being non-local).  Thus to branches I and II above
correspond two branches of twisted fields.  An analysis of the boundary
conditions of the fields in branch I, for instance, is exactly similar to
the one for $\C^2/\Z_{N,p}$ \cite{HKMM01} (modulo the change of $U(1)$ variables
mentioned before equation \ref{eq.action}).  The branch I twist fields actually
form a chiral ring isomorphic to that of $\C^2/\Z_{N,p}$, and is thus
{\it independent of $k$}.  In fact, these twist
fields are characterized by the same (left-moving) R-charges as the
fields in branch I, which are independent of $k$.  This means that the
ring doesn't change as we increase $k$, when the target space
$L_{N,p}\times \R_\phi$ approaches its large volume limit --
the flat space orbifold $\C^2/\Z_{N,p}$.  It is not surprising, therefore,
that branch I of our chiral ring in fact coincides with the chiral fields
of $\C^2/\Z_{N,p}$.  In fact, if the
 branch I fields are volume independent they should depend only on the
asymptotic structure of the target space; the background curvature is then
effectively replaced with a singularity in the same class of ALE spaces.

For branch II, which has only one generator, the analysis is similar to
that for $\C/\Z_{k}$ \cite{APS01}. These twist fields are similarly
characterized by the branch II (left-moving) R-charges, which
{\it do} depend on $k$.  These fields thus probe the high-curvature
region of $L_{N,p}\times \R_{dil}$, if only the preserved diagonal 
$U(1)$ symmetry of $SL(2,\R)/U(1) \times SU(2)/U(1)$.

\medskip

To clarify the structure of the complete chiral ring,  
the diagram of the corresponding R-charges, for the
theory with level $k=20$, orbifold order $N=10$ and discrete torsion
parameter $p=3$, is presented in Figure~1.
\begin{figure}
\begin{center}
\scalebox{.7}{\includegraphics{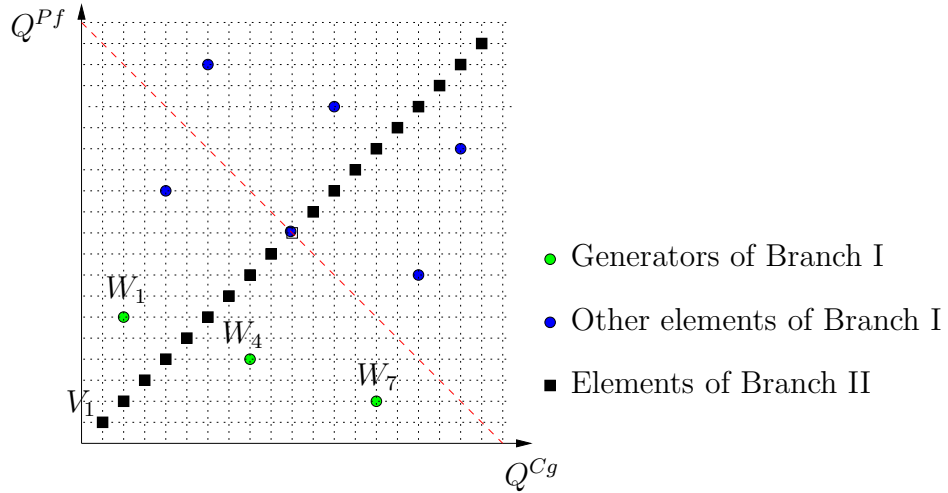}}
\put(-207,160){\makebox(0,0)[l]{$Q^{Pf}$}}
\put(-20,-10){\makebox(0,0)[l]{$Q^{Cg}$}}
\put(5,49){\makebox(0,0)[l]{Other elements of Branch I}}
\put(5,25){\makebox(0,0)[l]{Elements of Branch II}}
\put(5,73){\makebox(0,0)[l]{Generators of Branch I}}
\put(-171,60){\makebox(0,0)[l]{$W_1$}}
\put(-127,43){\makebox(0,0)[l]{$W_4$}}
\put(-76,27){\makebox(0,0)[l]{$W_7$}}
\put(-186,17){\makebox(0,0)[l]{$V_1$}}
\end{center}\label{fig.Rcharges}
\caption{{\small R-charges $Q^{pf}$ and $Q^{cg}$ of the twist fields
corresponding to the twisted sectors $W_i$, $V_i$ (\ref{eq.twistfields})
under the $\N=2$ superconformal algebras $SU(2)/U(1)_{k-2}$ and
$SL(2,\R)/U(1)_{k+2}$, respectively. Level $k=20$, orbifold 
order $N=10$ and discrete torsion parameter $p=3$. Grid in steps of $1/k$.}}
\end{figure}
Here we have used the fact that the R-charges are defined modulo 1.
The red line indicates the marginal fields, with $Q_{tot}=Q_{Pf}+Q_{cg}=1$.
The fields below this line are tachyons.  The diagram is in units of $1/k$,
and from it we can extract the structure of the ring -- in particular its
generators, as pictured.
In the general case, the analysis then follows directly from \cite{HKMM01}.
For completeness, we briefly review it here: for the branch I subring,
there are $[N,p]$ generators, where $[N,p]$ is the number of entries in
the continued fraction
\begin{equation}\label{eq.generatorsC2}
\frac{N}{p} = a_1-\frac{1}{a_2- \frac{1}{\cdots-\frac{1}{a_{[N,p]}}}} 
\equiv [a_1,\ldots,a_{[N,p]}]
\end{equation}
and the $a_i$ encode the ring structure of branch I.   In terms of the twisted
sectors (\ref{eq.twistfields}) this is
\begin{equation}
W_r^{a_r} = W_{(r+1)}W_{(r-1)}
\end{equation}
In the example of Figure~1, equation
(\ref{eq.generatorsC2}) becomes  $10/3=[4,2,2]$ and in particular there are
three generators.

On the other hand, branch II has only one generator, corresponding to the
twisted sector $V_1$, of order $k$ and R-charge $(p-1)/k$.  This branch
always includes tachyons, except in the supersymmetric case $p=1$ where
the whole branch collapses to the identity. 

\subsection{Tachyon condensation}

Having perturbed the theory with a chiral tachyon, one can study by a variety
of methods the chiral ring of the endpoint of the RG flow.  
For ALE spaces in the class
of $\C^2/\Z_{N,p}$, such as $\R_{dil} \times L_{N,p}$, the analysis in
\cite{HKMM01}\cite{Mar02} is very efficient,
so we may use it to study RG flows under our level-independent tachyons $W_r$.
It was shown there that the R-charge diagram provides an algebraic description
of the ALE structure of our space in terms of the pair $(N,p)$,
and that tachyon condensation here can be
thought of as blowing up particular curves of that space.  This method thus
identifies the ALE structure of the target space of the endpoint of 
condensation under our tachyon $W_r$
$$
(N,p) \longrightarrow_{W_r} (N_{r_1},p_{r_1})  
 \oplus \cdots \oplus (N_{r_t},p_{r_t}) 
$$
where the $r_i$ and $t$ depend on $W_r$.  Even though the 
ALE structure on the rhs
(ie. the chiral ring of the infrared theory) does not specify the 
worldsheet theory entirely, it is reasonable to
assume that the endpoint of the flow is in fact the
(set of) lens space with the corresponding ALE structure.
In particular, all the condensations determined in \cite{HKMM01} in this way
verified the $g_{cl}$ conjecture.

The branch II chiral subring (\ref{eq.twistfields}) coincides with the chiral
ring of the flat space orbifold $\C/\Z_{k}$ \cite{APS01}.  Naively then
(if one forgets about the branch II subring), one may be tempted to assume that
condensation of a $V_n$ tachyon will similarly drive the theory to
smaller deficit angles from the point of view of the twisted sectors:
$$
\C/\Z_{2k} \longrightarrow_{V_n} \C/\Z_{k_1(n)} \oplus \cdots \oplus \C/\Z_{k_t(n)}
$$
where in particular $\sum_i k_i = k$.  This would describe the gradual
disengagement of the $SL(2,\R)/U(1)$ and $SU(2)/U(1)$ factors, eventually 
leading to the direct product $SL(2,\R)/U(1)\times SU(2)/U(1)$.  
However, a more detailed analysis would be needed to investigate how 
such flows affect the entire chiral ring.

\section{Conclusion and Outlook}\label{sec.outlook}

We have embedded generalized lens spaces $L_{N,p}=SU(2)/\Z_{N,p}$ \cite{Bor03}
in string theory by adding
NS5-branes to the flat space orbifold $\C^2/\Z_{N,p}$ of \cite{HKMM01}.
This preserves the $\N=2$ superconformal worldsheet symmetry but leads
to spacetime tachyons.  The worldsheet
chiral ring can be divided into two parts: a level-independent part 
identical to that of $\C^2/\Z_{N,p}$, and another part identical to that
of $\C/\Z_{k}$.  Thus, chiral fields which are spacetime tachyons drive 
RG flows similar to those analysed in \cite{HKMM01} for these two cases.  

It would be important to explicitely write the partition functions of 
our backgrounds, to explore how modular invariant partition functions 
of non-rational theories exist in families parametrised by discrete 
parameters (such as the discrete torsion $p$) corresponding to different 
choices of twisted sectors.

It would also be interesting to have an effective description of
the tachyons in terms of NS5-branes interactions.  For instance, our 
branch I tachyons act only at the level of the orbifold, so here the 
question is shifted to how this orbifold is actually implemented in 
string theory.  To circumvent this particularity of our construction, 
one could try to embed a generalized lens space in
string theory by modifying a configuration which already includes a 
usual lens space $L_{N,1}$.     Examples include the appearance 
of Taub-NUT spaces as the 4d transverse space to KK-monopoles 
\cite{EH78}.  For $N$ monopoles distributed at positions $(\psi_i,\vec{x}_i)$
in transverse space, we get the multicentered Taub-NUT metric
\begin{equation}\label{taubmetric}
ds^2_{\perp} = V(x)^{-1} \left( d\psi + \omega \cdot x \right)^2 +
V(x) d\vec{x}\cdot\vec{x}
\end{equation}
where $\omega$ is a one-form determined by $V$ and we take 
the ansatz $V(x)=\epsilon + \sum_{i=1}^N |\vec{x}-\vec{x}_i|$.
Writing $\vec{x}$ in polar coordinates $(r,\theta,\phi)$ around an
$\vec{x}_i$, we can take the near horizon limit $r \rightarrow 0$.
The space spanned by $(\psi,\theta,\phi)$ is then a lens space $L_{M_i,1}$
where $M_i$ is the number of monopoles stacked at $\vec{x}_i$ (the
non-trivial $S^1$ is along $\psi$, whose periodicity assures the space is
smooth, turning the singularity at $x=x_i$ into a coordinate singularity).
It would be very interesting to find a modification of this configuration
such that the near horizon limit would include a generalized lens space
$L_{N,p}$ instead.

\paragraph{Acknowledgements:} It is a pleasure to acknowledge very useful
conversations with G. d'Appollonio, A. Recknagel, S. Ribault and also with
N. Lambert and D. Israel.  The author is supported by the grant
SFRH/BPD/18872/2004 provided by the Funda\c{c}\~ao para a
Ci\^encia e Tecnologia, Portugal.

\end{document}